\documentstyle[11pt,amssymb]{article}

\makeatletter
\@addtoreset{equation}{section}
\makeatother

\makeatletter
\def\cin#1{%
\edef\@tempa{\@ignspaftercomma,#1, \@end, }
\edef\@tempa{\expandafter\@ignendcommas\@tempa\@end}%
\if@filesw \immediate \write \@auxout {\string \citation {\@tempa}}\fi
\@tempcntb\m@ne \let\@h@ld\relax \def\@citea{}%
\@for \@citeb:=\@tempa\do {\@cmpresscites}%
\@h@ld}
\def\@ignspaftercomma#1, {\ifx\@end#1\@empty\else
   #1,\expandafter\@ignspaftercomma\fi}
\def\@ignendcommas,#1,\@end{#1}
\def\@cmpresscites{%
 \expandafter\let \expandafter\@B@citeB \csname b@\@citeb \endcsname
 \ifx\@B@citeB\relax 
    \@h@ld\@citea\@tempcntb\m@ne{\bf ?}%
    \@warning {Citation `\@citeb ' on page \thepage \space undefined}%
 \else
    \@tempcnta\@tempcntb \advance\@tempcnta\@ne
    \setbox\z@\hbox\bgroup 
    \ifnum0<0\@B@citeB \relax
       \egroup \@tempcntb\@B@citeB \relax
       \else \egroup \@tempcntb\m@ne \fi
    \ifnum\@tempcnta=\@tempcntb 
       \ifx\@h@ld\relax 
          \edef \@h@ld{\@citea\@B@citeB }%
       \else 
          \edef\@h@ld{\hbox{--}\penalty\@highpenalty
            \@B@citeB }%
       \fi
    \else   
       \@h@ld\@citea\@B@citeB
       \let\@h@ld\relax
 \fi\fi%
 \def\@citea{,\penalty\@highpenalty\hskip.13em plus.1em minus.1em}%
}
\def\@citex[#1]#2{\@cite{\cin{#2}}{#1}}%
\def\@cite#1#2{\leavevmode\unskip
  \ifnum\lastpenalty=\z@\penalty\@highpenalty\fi
  \ [{\multiply\@highpenalty 3 #1
      \if@tempswa,\penalty\@highpenalty\ #2\fi 
    }]\spacefactor\@m}
\makeatother

   \let\d=\delta 
    
  \let\n=\nu

\let\w=\omega

  \let\wp=\wedge     
\let\la=\label \let\ci=\cite \let\re=\ref
\let\se=\section   
\def\nn{\nonumber} \def\bd{
\begin{document}} \def\ed{\end{document}}
\def\ds{\documentstyle} \let\fr=\frac \let\bl=\bigl \let\br=\bigr
\let\Br=\Bigr \let\Bl=\Bigl 
\let\bm=\bibitem
\let\na=\nabla
\let\pa=\partial \let\ov=\overline 
\newcommand{\be}{\begin{equation}} 
\newcommand{\ee}{\end{equation}} 
\def\ba{\begin{array}}
\def\ea{\end{array}}
\def\ft#1#2{{\textstyle{{\scriptstyle #1}\over {\scriptstyle #2}}}}
\def\fft#1#2{{#1 \over #2}}
\def\del{\partial}
\def\vp{\varphi}
\def\sst#1{{\scriptscriptstyle #1}}
\def\oneone{\rlap 1\mkern4mu{\rm l}}
\def\simequiv{\buildrel\sim\over=}
\def\ep{\epsilon}
\def\td{\tilde}
\def\wtd{\widetilde}
\def\ie{{\it i.e.\ }}
\def\im{{\rm i}}
\def\dalemb#1#2{{\vbox{\hrule height .#2pt
        \hbox{\vrule width.#2pt height#1pt \kern#1pt
                \vrule width.#2pt}
        \hrule height.#2pt}}}
\def\square {\mathord{\dalemb{6.8}{7}\hbox{\hskip1pt}}}
\def\R {\rlap{\rm I}\mkern3mu{\rm R}}
\def\sR {\rlap{\hbox{$\scriptstyle\rm I$}}\mkern3mu{\hbox{$
\scriptstyle\rm R$}}}
\def\E {\rlap{\rm I}\mkern3mu{\rm E}}
\def\Z {\rlap{\sf Z}\mkern3mu{\sf Z}}
\def\0{{\sst{(0)}}}
\def\1{{\sst{(1)}}}
\def\2{{\sst{(2)}}}
\def\3{{\sst{(3)}}}
\def\4{{\sst{(4)}}}
\def\5{{\sst{(5)}}}
\def\6{{\sst{(6)}}}
\def\7{{\sst{(7)}}}
\def\8{{\sst{(8)}}}
\def\9{{\sst{(9)}}}
\def\ten {{\sst{(10)}}}
\def\n {{\sst{(n)}}}
\def\v {{\cal V}}
\def\semi {\ltimes}

\newcommand{\ho}[1]{$\, ^{#1}$}
\newcommand{\hoch}[1]{$\, ^{#1}$}
\newcommand{\bea}{\begin{eqnarray}} 
\newcommand{\eea}{\end{eqnarray}} 
\newcommand{\ra}{\rightarrow}
\newcommand{\lra}{\longrightarrow}
\newcommand{\Lra}{\Leftrightarrow}
\newcommand{\ap}{\alpha^\prime}
\newcommand{\bp}{\tilde \beta^\prime}
\newcommand{\tr}{{\rm tr} }
\newcommand{\Tr}{{\rm Tr} } 
\newcommand{\NP}{Nucl. Phys. }

\newcommand{\tamphys}{\it Center for Theoretical Physics,
Texas A\&M University, College Station, Texas 77843}

\newcommand{\auth}{I.V. Lavrinenko\hoch{\dagger}} 

\thispagestyle{empty}

\bd
\null
\vspace{-2cm}
\begin{flushright}
\hfill{CTP TAMU-33/99}\\
\hfill{hep-th/9909055}\\
\hfill{September 1999}\\
\end{flushright}

\begin{center}
{\bf{\large Bulk to Boundary Relation Between Topological Models\\
and Supergravity Theories}}

\vspace{15pt}
\auth

\vspace{10pt}
{\tamphys}

\vspace{15pt}
\underline{ABSTRACT}
\end{center}

We establish a direct correspondence between certain higher-rank
$p$-form Chern-Simons topological type theories in the bulk of a
manifold with boundary and particular sectors of supergravity models
on the boundary, provided that certain boundary conditions are
satisfied.  The cases we investigate include eleven-dimensional
supergravity and both of the type II theories in ten dimensions.

{\vfill\leftline{}\vfill
\vskip  10pt

\footnoterule
{\footnotesize \hoch{\dagger} e-mail: 
lavrik{@}rainbow.physics.tamu.edu\vskip -12pt} \vskip 10pt

\pagebreak

\se{Introduction}
\setcounter{page}{1}

Superstring models and their low-energy supergravity limits are  among the
most
complex and exciting of physical theories, and they are also among the
most
promising as far as phenomenological expectations are concerned. Even
though so
far no concrete predictions have been made, the intrinsic mathematical
consistency and beauty of these theoretical constructions have  been
attracting
much attention. In this paper we will uncover one more rather interesting
aspect of the mathematical conspiracies that make these theories so
unique. It
is well known that symmetries play an important role in all physical
theories. 
It is a simple observation that if one is able to devise a formulation
where
these symmetry properties are manifest, then previously hidden  aspects of
the
theory can become more transparent. 

   In this paper we shall demonstrate that there is a direct relation
between
certain unusual Chern-Simons (CS) type theories in higher-dimensional
spacetimes with boundaries and particular sectors of supergravity models
on the
boundary of the spacetime.  In order to make this relation explicit shall
develop a certain first-order formalism.  An almost identical approach,
called
the doubled-field formalism, has been used in the past \ci{cjlp2} to unify
the
duality symmetries and gauge invariances of supergravity theories, and
later
on  the formalism has been successfully applied to obtain certain, rather
general, relations among $p$-brane tensions \ci{llps}. However, to prevent
a
confusion, we should emphasize that the approach used in this paper is
{\it
not} identical to that introduced in \cite{cjlp2}. For instance, our
definitions for the field-strengths are different, and all fields in the
theory
are put on an equal footing, including the dilatons.  As for the general
structure, it is strikingly reminiscent of the situation in an ordinary
CS-theory in a three-dimensional spacetime with boundary, where it is well
known that in the bulk the theory is topological, and all the dynamics
takes
place only on the boundary, where the degrees of freedom are described by
the
chiral Wess-Zumino-Witten-Novikov (WZWN) model \ci{cswzwn}.  However, now
the
CS-theories are replaced by some higher $p$-form analogs, and instead of a
WZWN-model we obtain various supergravity theories. 

    Boundary conditions play a rather important role in the construction
\ci{b_b, b_b1, b_b2, b_b3}.  One needs to impose these conditions in order
for
the variational principle to make sense, and to maintain gauge invariance
in
the theory on a manifold with boundary. The CS-type theories we obtain
have
much in common with the so-called higher-rank BF theories \ci{bal2, fpr,
ns},
although in our approach these theories acquire a non-Abelian
generalization.
Even though many issues have been left unaddressed, the results obtained
may
indicate that there are hidden topological sectors in supergravity
theories, or
even that these theories may be described by pure topological models on
manifolds with boundaries. It is also conceivable that the topological
theories
we describe may have applications of their own, related to some properties
of
higher-dimensional manifolds {\it without} boundaries \ci{thompson}.
However,
the fact that these models preserve metric-independence after quantization
still remains to be proved.

\se{D=11 Supergravity}   

In this section we shall consider the example of $D=11$ supergravity. It
is a
simplest case possible, however it already includes many characteristic
features. We would like to start with the following Lagrangian in $D=12$ 
\be  {\cal L}=\ft12(A_\4\wp dA_\7+A_\7\wp dA_\4)+ \ft16 A_\4\wp A_\4\wp
A_\4,\la{lagr}  \ee   
where $A_\4$ and $A_\7$ are four-form and
seven-form gauge fields.

It is easy to see that this Lagrangian changes by total derivative
under the following gauge transformations
\bea
\d A_\4 &=& d\w_\3,\la{g_tr}\\
\d A_\7 &=& -\w_\3\wp A_\4+d\w_\6,\nn
\eea
where $\w_\3$ and $\w_\6$ are arbitrary three-forms and six-forms.

In fact
\be
\d {\cal L}=\ft12 d(d\w_\3\wp A_\7-d\w_\6\wp A_\4).
\ee
The following brackets hold
\be
\{\d(\w^1_\3), \d(\w^2_\3)\} = 
\d(\w^2_\3\wp\w^1_\3), \qquad
\{\d(\w^1_\3), \d(\w^2_\6)\} = 0
\ee
If the manifold has no boundary, the action obtained by integrating this
Lagrangian over the whole spacetime is invariant with respect to these
gauge
transformations.  In fact the Lagrangian (\re{lagr}) can be obtained by
following the canonical procedure for CS-theories. Let us define the
following
field-strengths
\bea
F_\5 &=& dA_\4,\la{fields}\\
F_\8 &=& dA_\7+\ft12 A_\4\wp A_\4\nn.
\eea
It can be checked that under the gauge transformations (\re{g_tr}) fields
(\re{fields}) transform covariantly
\bea
\d F_\5 &=& 0,\\
\d F_\8 &=& \w_\3\wp F_\5.\nn
\eea
Out of these field-strengths one can construct the analog of the Chern
class
\be
F_\5\wp F_\8.\la{class}
\ee
It is easy to see that the class (\re{class}) is invariant with respect to
the
gauge transformations (\re{g_tr}), and also that the following identity
holds
\be
F_\5\wp F_\8 = d(\ft12(A_\4\wp dA_\7+A_\7\wp dA_\4)+
\ft16 A_\4\wp A_\4\wp A_\4) = d{\cal L}.
\ee
   Thus everything is very reminiscent of ordinary Chern-Simons theory,
and
what we have said so far, with the exception of gauge invariance, is valid
whether we have a boundary or not. In the presence of a boundary, things
change
in a rather interesting way. First of all, it seems that the theory is no
longer gauge invariant, due to the total derivative which, upon
integration,
gives rise to a non-vanishing term on the boundary.  However, we know how
to
deal with this problem.  The solution again comes from the conventional
CS-theory, where the same problem arises; one imposes some boundary
conditions
in order to restore gauge invariance \ci{b_b, b_b1, b_b2, b_b3}. The
trivial
condition that the gauge transformations vanish on the boundary is not
interesting, since it appears to be too strong. The only other choice is
to
impose the following relations
\bea
d\w_\3 = {{*d}\w_\6} |_{\del M_{12}},\la{b_cond}\\
A_\7 = {{*A}_\4} |_{\del M_{12}},\nn
\eea
where the Hodge $*$ is taken with respect to the eleven-dimensional metric
on
the boundary. If the relations (\ref{b_cond}) hold then the boundary term
vanishes, and gauge invariance is restored. It can also be checked that
the
variational principle is well defined in this case. Normally, if there is
a
boundary and the Lagrangian changes by a total derivative, problems may
arise
owing to the fact that the functional derivative is not well defined
unless one
adds some extra terms on the boundary to cancel terms coming from the
bulk. In
our situation the boundary contribution from the bulk has the form
\be
\int\d{\cal L}=\ft12\int_{\del M_{12}} 
(A_\4\wp\d A_\7-A_\7\wp\d A_\4),
\ee
and it vanishes if boundary conditions (\re{b_cond}) are satisfied.

This is not the end of story though.  The action of twelve-dimensional
theory
(\re{lagr}) can be written as
\bea
{\cal S} &=& \int_{\del M_{12}}\ft12 A_\4\wp A_\7+
\int_{M_{12}} (A_\7\wp dA_\4+\ft16 A_\4\wp A_\4\wp A_\4)\\
&=& \int_{\del M_{12}}\ft12 A_\4\wp {*A}_\4+
\int_{M_{12}} (A_\7\wp dA_\4+\ft16 A_\4\wp A_\4\wp A_\4),\nn
\eea
where in the second line we used the boundary conditions introduced in
(\ref{b_cond}).

  Now, variation of the $A_\7$-field (which simply plays the role of a
constraint) in the bulk gives the condition $dA_\4=0$, which implies that
(at
least locally) $A_\4=dA_\3$, and by continuation the same must be true on
the
boundary.  Thus if one substitutes this relation back into the Lagrangian
the
result is a theory on the boundary whose Lagrangian is identical to the
one for
$D=11$ supergravity (without Einstein-Hilbert term).  There are also some
subtleties with the gauge symmetry of the theory. The generators are
simply
constraints coming from the variation over the components of $A_\4$ and
$A_\7$
gauge fields with a time index, since these have no conjugate momenta.
They are
\be
G_m(\w_\6) = \int_{M_{11}}\w_\6\wp dA_\4, \qquad
G_e(\w_\3) = \int_{M_{11}}\w_\3\wp (dA_\7+\ft12 A_\4\wp A_\4),
\ee
where $\w_\3$ and $\w_\6$ are arbitrary three-forms and six-forms, and
the integration is over only the spatial coordinates.

All states in the theory must satisfy the constraints $G_e |\ \rangle =
0$, and
$G_m |\ \rangle = 0$.  If there is no boundary, the algebra of these
constraints can be easily calculated using the fact that the $A_\4$ and
$A_\7$
fields are canonically conjugate to each other.  The algebra gives rise to
the
following brackets
\be
\{ G_e(\w^1_\3), G_e(\w^2_\3) \} = G_m(\w^1_\3\wp\w^2_\3),
\qquad \{ G_e(\w^1_\3), G_m(\w^2_\6) \} = 0.
\ee
However, if the spacetime has a boundary, one runs into a problem.
Functional
derivatives of the constraints are not well defined, and one needs to add
some
boundary terms to the generators to cancel these contributions, and these
extra
terms will play the role of boundary symmetries. Namely one needs to add
$g_m =
-\int_{\del M_{11}} \w_\6\wp A_\4$ to $G_m$, and $g_e = -\int_{\del
M_{11}}
\w_\3\wp A_\7 = -\int_{\del M_{11}}\w_\3\wp *A_\4$ to $G_e$, and also the
commutator between the (modified) $G_e$ and $G_m$ generators changes,
becoming
$\{ G_e(\w_\3), G_m(\w_\6) \} = \int_{\del M_{11}}\w_\3\wp d\w_\6$.  After
we
have imposed the constraints $G_e |\ \rangle = 0$, and $G_m |\ \rangle =
0$ the
gauge symmetry in the bulk therefore gives rise to the symmetry of a
theory on
the boundary
\be
\{ g_e(\w^1_\3), g_e(\w^2_\3) \} = 
g_m(\w^1_\3\wp\w^2_\3),\qquad
\{ g_e(\w_\3), g_m(\w_\6) \} = 
\int_{\del M_{11}}\w_\3\wp d\w_\6.
\ee
   A similar idea has been employed in CS-theory in $D=3$, where it
gave rise to certain conformal models on the boundary with affine
symmetries \ci{b_b1, b_b2, b_b3}.
 
As a {\it nontrivial} consistency check one can verify that
Jacobi-identity is satisfied. The only interesting one is

\pagebreak

\bea
\{ g_e(\w^1_\3), \{ g_e(\w^2_\3), g_e(\w^3_\3) \} \} +
\{ g_e(\w^3_\3), \{ g_e(\w^1_\3), g_e(\w^2_\3) \} \}+\\ 
\{ g_e(\w^2_\3), \{ g_e(\w^3_\3), g_e(\w^1_\3) \} \} = \nn\\
\int_{\del M_{11}} (\w^1_\3\wp d(\w^2_\3\wp\w^3_\3)+
\w^2_\3\wp d(\w^3_\3\wp\w^1_\3)+
\w^3_\3\wp d(\w^1_\3\wp\w^2_\3)) = \nn\\
-2 \int_{\del M_{11}} d(\w^1_\3\wp\w^2_\3\wp\w^3_\3) = 0.\nn
\eea
\se{Type IIA-Theory in D=10}

Now we would like to consider the case of the Type IIA theory in
$D=10$. Here things are sightly more complicated, owing to the increase
in the number of fields, but main ideas stay the same. The ten-dimensional
Lagrangian has the form
\bea
{\cal L}_{10} &=& R {*\oneone} -\ft12 {{*d}\phi}\wp d\phi - \ft12
e^{\fft32\phi}\, {{*{\cal F}}_\2}\wp {\cal F}_\2 -
\ft12 e^{-\phi}\,{{*F}_\3}\wp F_\3\nn\\
&& - \ft12 e^{\fft12\phi}\, {{*F}_\4}\wp F_\4 -
\ft12 dA_\3\wp dA_\3\wp A_\2\ ,
\eea
where $F_\4=dA_\3 -dA_\2\wp {\cal A}_\1$, $F_\3=dA_\2$ and ${\cal
F}_\2= d{\cal A}_\1$. From this, it follows that the equations of motion
for the antisymmetric tensor and scalar fields are:
\bea
d(e^{\fft12\phi} \, {{*F}_\4}) &=& -F_\4\wp F_\3\ .\nn\\
d(e^{-\phi}\, {{*F}_\3}) &=&-{\cal F}_\2\wp (e^{\fft12\phi}{{*F}_\4}) -
\ft12 F_\4\wp F_\4\ ,\la{2a}\\
d(e^{\fft32\phi}\, {{*{\cal F}}_\2}) &=& -F_\3\wp (e^{\fft12\phi}\, 
{{*F}_\4}) \ ,\nn\\
d{{*d}\phi} &=& -\ft14 F_\4\wp ( e^{\fft12\phi}\,{{*F}_\4}) -
\ft12 F_\3\wp (e^{-\phi}\,{{*F}_\3}) -
\ft34 {\cal F}_\2\wp (e^{\fft32\phi}\, {{*{\cal F}}_\2}) \ .\nn
\eea
Let us make the following redefinitions
\be
e^{\fft14\phi} F_\4 = B_\4,\qquad  e^{-\fft12\phi} F_\3 = B_\3,\qquad
e^{\fft34\phi} {\cal F}_\2 = B_\2,\qquad d\phi = B_\1,
\ee
and introduce the following new fields
\be
B_\6 = {*B}_\4,\qquad B_\7 = {*B}_\3,\qquad B_\8 = {*B}_\2,\qquad B_\9 =
{*B}_\1.\la{2a_dual}
\ee
In terms of these new fields equations (\re{2a}) can be written as follows
\bea
H_\2 &=& dB_\1 = 0,\la{2a_cs}\\
H_\3 &=& dB_\2-\ft34 B_\1\wp B_\2 = 0,\nn\\ 
H_\4 &=& dB_\3+\ft12 B_\1\wp B_\3 = 0,\nn\\
H_\5 &=& dB_\4-\ft14 B_\1\wp B_\4-B_\3\wp B_\2 = 0,\nn\\
H_\7 &=& dB_\6+\ft14 B_\1\wp B_\6+B_\4\wp B_\3 = 0,\nn\\
H_\8 &=& dB_\7-\ft12 B_\1\wp B_\7+B_\2\wp B_\6+\ft12 B_\4\wp B_\4 =
0,\nn\\
H_\9 &=& dB_\8+\ft34 B_\1\wp B_\8+B_\3\wp B_\6 = 0,\nn\\
H_\ten &=& 
dB_\9+\ft14 B_\4\wp B_\6+\ft12 B_\3\wp B_\7+\ft34 B_\2\wp B_\8 = 0,\nn
\eea
where we have also introduced new field-strengths $H_\2$, $H_\3$, $H_\4$, 
$H_\5$, $H_\7$, $H_\8$, $H_\9$, and $H_\ten$.

At this stage we would like to emphasize that there is a {\it one-to-one}
correspondence between original equations of motion (\re{2a}) and the
system
(\re{2a_cs}), provided that relations (\re{2a_dual}) are satisfied.
Indeed, one
can solve for $B_\1$ first (locally), and then for $B_\2$, $B_\3$, and
$B_\4$,
and after that, using relations (\re{2a_dual}) for $B_\6$, $B_\7$, $B_\8$,
and
$B_\9$, so eventually one recovers the original system (\re{2a}).

It turns out that the system (\re{2a_cs}) is invariant under the
following gauge transformations
\bea
\d B_\1 &=& d\w,\la{2a_symmetry}\\
\d B_\2 &=& d\w_\1+\ft34\w_\1\wp B_\1+\ft34\w B_\2,\nn\\
\d B_\3 &=& d\w_\2+\ft12\w_\2\wp B_\1-\ft12\w B_\3,\nn\\
\d B_\4 &=& d\w_\3+\ft14\w_\3\wp B_\1+\w_\2\wp B_\2+
\w_\1\wp B_\3+\ft12\w B_\4,\nn\\
\d B_\6 &=& d\w_5-\ft14\w_5\wp B_\1-\w_\3\wp B_\3-
\w_\2\wp B_\4-\ft14\w B_\6,\nn\\
\d B_\7 &=& d\w_\6-\ft12\w_\6\wp B_\1-\w_5\wp B_\2-
\w_\3\wp B_\4-\w_\1\wp B_\6+\ft12\w B_\7,\nn\\
\d B_\8 &=& d\w_\7-\ft34\w_\7\wp B_\1-\w_5\wp B_\3-
\w_\2\wp B_\6-\ft34\w B_\8,\nn\\
\d B_\9 &=& d\w_\8-\ft34\w_\7\wp B_\2+\ft12\w_\6\wp B_\3-
\ft14\w_5\wp B_\4\nn\\
&-&\ft14\w_\3\wp B_\6-\ft12\w_\2\wp B_\7-\ft34\w_\1\wp B_\8,\nn
\eea
where $\w$, $\w_\1$, $\w_\2$, $\w_\3$, $\w_5$, $\w_\6$, $\w_\7$, and
$\w_\8$ are arbitrary $p$-forms of the appropriate degree.  This is
actually an infinitesimal form of the transformations, but it is good
enough for our purposes. Another way to say that transformations
(\re{2a_symmetry}) leave the system (\re{2a_cs}) invariant is to claim
that field-strengths $H_i$ transform covariantly, namely that they
transform
into each other, which can be checked by a direct calculation. Now we
would like to apply the same idea as we did in $D=11$ supergravity to the
type
IIA theory. Let us consider the following Lagrangian in
eleven-dimensional spacetime

\pagebreak

\bea
{\cal L} &=& \ft12(B_\6\wp dB_\4-B_\4\wp dB_\6)-
\ft12(B_\3\wp dB_\7+B_\7\wp dB_\3)\la{d112a}\\
&+& \ft12(B_\8\wp dB_\2-B_\2\wp dB_\8)-
\ft12(B_\1\wp dB_\9+B_\1\wp\ dB_\9)\nn\\
&-& \ft14 B_\6\wp B_\1\wp B_\4-B_\6\wp B_\3\wp B_\2-
\ft12 B_\7\wp B_\1\wp B_\3\nn\\
&-& \ft34 B_\8\wp B_\1\wp B_\2-\ft12 B_\4\wp B_\4\wp B_\3.\nn 
\eea
As a nice consistency check, one can derive the following formal identity
\be
d{\cal L} = H_\7\wp H_\5-H_\ten\wp H_\2-H_\8\wp H_\4,
\ee
where we are using the definitions (\re{2a_cs}). Again it is evident that
the Lagrangian (\re{d112a}) can be obtained from the descent process
from a thirteen-dimensional spacetime.

First of all under the gauge transformations (\re{2a_symmetry}) the
Lagrangian
(\re{d112a}) transforms as a total derivative; for example the $\w$ and
$\w_\8$ transformations give rise to the following term
\be
\d {\cal L} = \ft12 d(d\w_\8\wp B_\1+d\w B_\9).
\ee
All other transformations produce similar results, namely it is always a
sum
of two terms each of which is a wedge product of the derivative of a
gauge parameter and a field of complementary degree. 
 
   Therefore if there is no boundary in the eleven-dimensional spacetime
we have gauge invariance; otherwise it is lost. Of course, we can
restore gauge invariance by imposing appropriate boundary conditions
on the fields and gauge parameters.  Not too surprisingly, they turn out
to be the same as equations (\re{2a_dual}), plus corresponding
conditions for the gauge-parameters, namely $d\w_\8 = {*d}\w$, $d\w_\7 =
{*d}\w_\1$, and so on. Also following the procedure described in the
second
section, namely through integrating certain terms by parts, we get
the action in eleven-dimensional bulk plus boundary terms
\bea
{\cal S} &=& -\ft12\int_{\del M_{11}} ({*B}_\4\wp B_\4+{*B}_\3\wp B_\3+
{*B}_\2\wp B_\2+{*B}_\1\wp B_\1)\la{d11_2a}\\
&+& \int_{M_{11}} (B_\6\wp dB_\4-B_\7\wp dB_\3+B_\8\wp dB_\2-
B_\9\wp dB_\1\nn\\
&-& \ft14 B_\6\wp B_\1\wp B_\4-B_\6\wp B_\3\wp B_\2-
\ft12 B_\7\wp B_\1\wp B_\3\nn\\
&-& \ft34 B_\8\wp B_\1\wp B_\2-\ft12 B_\4\wp B_\4\wp B_\3).\nn 
\eea 
Now, let us evaluate the equations of motion for $B_\9$, $B_\8$, $B_\7$,
and
$B_\6$ in the bulk, which imply that the rest of the fields are pure gauge
(everywhere, including the boundary).  After integrating some terms by
parts,
we obtain a Lagrangian on the boundary identical to the one for the type
IIA
theory in $D=10$, where pure gauge degrees of freedom from the bulk become
dynamical.  We should  remark that presumably this procedure can be
carried out
even at the quantum (functional integral) level, since the $B_\9$, $B_\8$,
$B_\7$, and $B_\6$ fields simply play the role of Lagrange multipliers.
Thus
the integrations over these fields produce delta-functions which impose
that
the $B_\1$, $B_\2$, $B_\3$, and $B_\4$ gauge fields are flat, or pure
gauge. 
The associated redefinitions of variables may give rise to non-trivial
Jacobians in the functional integral measure, and in turn may produce some
dynamics for the metric on the boundary. It is very tempting to conjecture
that
these Jacobians will eventually produce an Einstein-Hilbert curvature term
on
the boundary, and that the theory we are describing is all that is needed
to
produce the complete supergravity action on the boundary, including
gravity. 
Clearly these issues require a more careful investigation.

\se{Type IIB-Theory in D=10}

  There is no covariant Lagrangian for type IIB supergravity, since
it includes a self-dual 5-form field strength.  However one can write
down covariant equations of motion.  In order to make
manifest their global $SL(2,\R)$ symmetry, it is useful first to
assemble the dilaton $\phi$ and axion $\chi$ into a $2\times 2$
matrix:
\be
{\cal M} = \pmatrix{ e^\phi & \chi\, e^\phi\cr
                     \chi\, e^\phi & e^{-\phi} + \chi^2\, e^\phi}
\ee
Also, define the $SL(2,\R)$-invariant matrix
\be
\Xi = \pmatrix{ 0 & 1\cr
                   -1 & 0}\ ,
\ee
and the two-component column vector of 2-form potentials
\be
A_\2 = \pmatrix{A_\2^1 \cr A_\2^2}\ .
\ee
Here $A_\2^1$ is the R-R potential, and $A_\2^2$ is the NS-NS
potential.  The bosonic matter equations of motion can then be written as
\bea
d{{*H}_\5} &=& -\ft12 \ep_{ij}\, F_\3^i\wp F_\3^j\ ,\nn\\
d({\cal M} {{*H}_\3}) &=& H_\5 \wp \Xi\, H_\3\ ,\nn\\
d(e^{2\phi}\, {{*d}\chi}) &=& e^{\phi}\, F_\3^2\wp   
{{*F}_\3^1}\ ,\nn\\
d{{*d}\phi} &=& e^{2\phi}\, d\chi\wp {{*d}\chi} +\ft12 e^{\phi}\,
F_\3^1\wp {{*F}_\3^1} -\ft12 e^{-\phi}\, F_\3^2\wp {{*F}_\3^2}\ ,
\la{2beom}
\eea
where $F_\3^1 = dA_\2^1 -\chi\, dA_\2^2$, $F_\3^2=dA_\2^2$,
$H_\3=dA_\2$, and $H_\5 = dB_\4 - \ft12 \ep_{ij}\, A_\2^i\wp dA_\2^j$.

Now let us define the following new fields
\bea
C^1_\1 &=& d\phi,\qquad C^2_\1 = e^{\phi} d\chi,\qquad 
C_\5 = H_\5\la{new_2b}\\
C^1_\3 &=& e^{\fft12\phi} dA^1_\2+\chi e^{\fft12\phi} dA^2_\2,\qquad
C^2_\3 = e^{-\fft12\phi} dA^2_\2,\nn\\
\eea
and 
\bea
C^1_\7 &=& {*C}^1_\3,\qquad C^2_\7 = {*C}^2_\3,\la{2b_dual}\\
C^1_\9 &=& {*C}^1_\1,\qquad C^2_\9 = {*C}^2_\1,\nn
\eea
where we are also assuming that $C_\5 = {*C}_\5$

It turns out that in terms of these fields the system (\re{2beom}) can be
written as follows
\bea
G^1_\2 &=& dC^1_\1 = 0, \la{2b_cs}\\
G^2_\2 &=& dC^2_\1-C^1_\1\wp C^2_\1 = 0,\nn\\
G_\6 &=& dC_\5-C^2_\3\wp C^1_\3 = 0,\nn\\
G^1_\4 &=& dC^1_\3-\ft12 C^1_\1\wp C^1_\3-C^2_\1\wp C^2_\3 = 0,\nn\\
G^2_\4 &=& dC^2_\3+\ft12 C^1_\1\wp C^2_\3 = 0,\nn\\
G^1_\8 &=& dC^1_\7+\ft12 C^1_\1\wp C^1_\7-C_\5\wp C^2_\3 = 0,\nn\\
G^2_\8 &=& dC^2_\7-\ft12 C^1_\1\wp C^2_\7+C^2_\1\wp C^1_\7+
C_\5\wp C^1_\3 = 0,\nn\\
G^1_\ten &=& dC^1_\9-C^2_\1\wp C^2_\9-\ft12 C^1_\3\wp C^1_\7+
\ft12 C^2_\3\wp C^2_\7 = 0,\nn\\
G^2_\ten &=& dC^2_\9+C^1_\1\wp C^2_\9-C^2_\3\wp C^1_\7 = 0,\nn
\eea
where we have also introduced corresponding field-strengths $G^i_j$.

Again we would like to emphasize that the system (\re{2b_cs}) is 
{\it completely} equivalent to the original set of equations (\re{2beom}) 
provided that conditions (\re{2b_dual}) are satisfied.

One can easily check that this new system is invariant with respect to
the following gauge transformations
\bea
\d C^1_\1 &=& d\w^1,\la{2b_gtr}\\
\d C^2_\1 &=& d\w^2-C^1_\1\w^2+\w^1 C^2_\1,\nn\\
\d C^1_\3 &=& d\w^1_\2-\ft12 C^1_\1\wp \w^1_\2+\ft12\w^1 C^1_\3+
\w^2 C^2_\3-\w^2_\2\wp C^2_\1,\nn\\
\d C^2_\3 &=& d\w^2_\2+\ft12 C^1_\1\wp\w^2_\2-\ft12\w^1 C^2_\3,\nn\\
\d C_\5 &=& d\w_\4-\w^1_\2\wp C^2_\3+\w^2_\2\wp C^1_\3,\nn\\
\d C^1_\7 &=& d\w^1_\6+\ft12 C^1_\1\wp\w^1_\6-\ft12\w^1 C^1_\7-
\w^2_\2\wp C_\5,\nn\\
\d C^2_\7 &=& d\w^2_\6-\ft12 C^1_\1\wp\w^2_\6+\ft12\w^1 C^2_\7-
\w^2 C^1_\7+\w^1_\2\wp C_\5+\w^1_\6\wp C^2_\1,\nn\\
\d C^1_\9 &=& d\w^1_\8+\ft12\w^2_\6\wp C^2_\3-\ft12\w^1_\6\wp C^1_\3-
\ft12\w^2_\2\wp C^2_\7-\w^2_\8\wp C^2_\1+\w^2 C^2_\9,\nn\\
\d C^2_\9 &=& d\w^2_\8+C^1_\1\wp\w^2_\8-\w^1_\6\wp C^2_\3+
\w^2_\2\wp C^1_\7-\w^2 C^2_\9,\nn
\eea
where $\w^i_j$ are arbitrary $p$-form parameters.

Now, as always, we go one dimension higher, and introduce the following
Lagrangian in eleven-dimensional spacetime
\bea
{\cal L} = \ft12(C^1_\1\wp dC^1_\9+C^1_\9\wp dC^1_\1)+
\ft12(C^2_\1\wp dC^2_\9+C^2_\9\wp dC^2_\1)+\la{2b_cs_lagr}\\
\ft12(C^1_\3\wp dC^1_\7+C^1_\7\wp dC^1_\3)+
\ft12(C^2_\3\wp dC^2_\7+C^2_\7\wp dC^2_\3)+\nn\\
\ft12 C_\5\wp dC_\5+
C^1_\1\wp C^2_\9\wp C^2_\1+\ft12 C^1_\3\wp C^1_\1\wp C^1_\7+\nn\\
\ft12 C^1_\1\wp C^2_\3\wp C^2_\7+C^2_\3\wp C^2_\1\wp C^1_\7+
C_\5\wp C^1_\3\wp C^2_\3.\nn
\eea
Now one has to go through the same sequence of steps, and observe that
under the symmetry (\re{2b_gtr}) this Lagrangian changes by a total
derivative, so that if there is a boundary then gauge invariance is
broken. To restore the symmetry one needs to impose some boundary
conditions, and of course, not too surprisingly, they turn out to be
the same as equations (\re{2b_dual}), {\it including the self-duality
constraint for 5-form field-strength}, plus the similar relations for
gauge-parameters.  Let us just demonstrate how this comes about for the
5-form. Under the symmetry $\d C_\5 = d\w_\4$ the action transforms as
follows
\be
\d{\cal S} = \int_{M_{11}}\d{\cal L} = 
\ft12\int_{\del M_{11}} d\w_\4\wp C_\5
\ee
Now if we impose the conditions $C_\5 = {*C}_\5$ and $d\w_\4 = {*d}\w_\4$
then
remarkably this term vanishes, by virtue of the Lorentzian metric
signature. 
Again one can easily check that the theory lives on the boundary only,
where it
coincides with $D=10$ type IIB-Theory (without gravity, of course). In
order to
demonstrate this one needs to write out the action for the Lagrangian
(\re{2b_cs_lagr}) in such a form that the bulk part does not have terms
where
the $C^i_9$ and $C^i_7$ fields are covered by derivatives.  They will
therefore
play the role of bulk Lagrange multipliers which impose certain
constraints on
the rest of the fields, plus boundary terms with appropriate boundary
conditions to have gauge invariance.  After integrating out the $C^i_9$
and
$C^i_7$ fields in the bulk and solving constraints they impose, and
integrating
certain terms by parts, one derives the type IIB action together with the
extra
equations for $C_\5$.

\se*{Acknowledgements}

I am grateful to Chris Pope for useful discussions and constant
encouragement.

\ed